\documentclass[preprint,pra,aps,showpacs,preprintnumbers,amsmath,amssymb,eqsecnum]{revtex4}

\usepackage{graphicx}
\usepackage{dcolumn}
\usepackage{bm}
\usepackage{verbatim}
\usepackage{epsfig}
\usepackage{epstopdf}

\begin{document}


\title{Exact Phase Space Localized Projectors from Energy Eigenstates}

\author{J.J.Halliwell}%
\affiliation{Blackett Laboratory \\ Imperial College \\ London SW7
2BZ \\ UK }



\begin{abstract}
In investigations of the emergence of classicality from quantum
theory, a useful step is the construction of quantum operators
corresponding to the classical notion that the system resides in a
region of phase space. The simplest such constructions, using
coherent states, yield operators which are {\it approximate}
projection operators -- their eigenvalues are approximately equal
to $1$ or $0$. Such projections may be shown to have close to
classical behaviour under time evolution and these results have
been used to prove some useful results about emergent classicality
in the decoherent histories approach to quantum theory. Here, we
show how to use the eigenstates of a suitably chosen Hamiltonian
to construct {\it exact} projection operators which are localized
on regions of phase. We elucidate the properties of such operators
and explore their time evolution. For the special case of the
harmonic oscillator, the time evolution is particularly simple,
and we find sets of phase space localized histories which are
exactly decoherent for any initial state and have probability $1$
for classical evolution. These results show how
approximate decoherence of histories and classical predictability
for phase space histories may be made exact in certain cases.

\end{abstract}

\pacs{03.65.-w, 03.65.Yz, 03.65.Ta}


\maketitle

\newcommand\beq{\begin{equation}}
\newcommand\eeq{\end{equation}}
\newcommand\bea{\begin{eqnarray}}
\newcommand\eea{\end{eqnarray}}

\def\A{{\cal A}}
\def\D{\Delta}
\def\H{{\cal H}}
\def\E{{\cal E}}
\def\p{\partial}
\def\la{\langle}
\def\ra{\rangle}
\def\ria{\rightarrow}
\def\x{{\bf x}}
\def\y{{\bf y}}
\def\k{{\bf k}}
\def\q{{\bf q}}
\def\p{{\bf p}}
\def\P{{\bf P}}
\def\r{{\bf r}}
\def\s{{\sigma}}
\def\a{\alpha}
\def\b{\beta}
\def\e{\epsilon}
\def\U{\Upsilon}
\def\G{\Gamma}
\def\om{{\omega}}
\def\Tr{{\rm Tr}}
\def\ih{{ \frac {i} { \hbar} }}
\def\trho{{\rho}}

\def\au{{\underline \alpha}}
\def\bu{{\underline \beta}}
\def\pp{{\prime\prime}}
\def\id{{1 \!\! 1 }}
\def\half{\frac {1} {2}}

\def\jjh{j.halliwell@ic.ac.uk}

\section{Introduction}

The question of the how classical mechanics emerges from quantum
mechanics continues to attract much interest \cite{Hal1}. There are
are many different aspects to this question, but this paper
concerns one specific aspect, namely, the question of
characterizing phase space localization in quantum mechanics and
the time evolution of phase space localized states.

\begin{figure}[h]
\begin{center}
\includegraphics[width=5in]{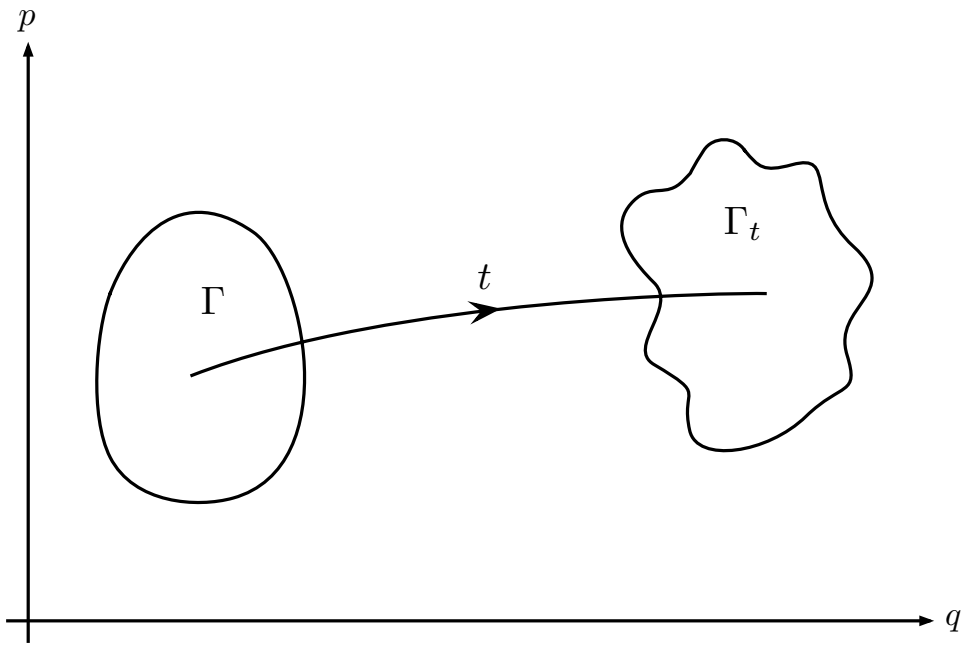}
\caption{The classical evolution of a phase space cell $\Gamma$. }
\label{Fig1}
\end{center}
\end{figure}

In classical mechanics, we may construct a phase space probability
distribution function $ w(p,q)$ that is perfectly localized in a
phase space cell $\Gamma$. The time evolution of such a
probability distribution is straightforward -- it evolves to
another distribution perfectly localized about a cell $\Gamma_t$,
the evolution of the original cell along the classical orbits. (See Fig.(\ref{Fig1})).
In fact, the existence and evolution properties of such distribution
functions characterize classical behaviour. In looking for the
emergence of classical behaviour from quantum theory a reasonable
approach is to attempt to reproduce these steps in the quantum
theory as closely as possible.

Of course in quantum theory we do not expect to be able
to reproduce these properties exactly. First of all, the fact that
positions and momenta are represented by non-commuting operators
means that it will not be possible to find states that are
perfectly localized in phase space. Secondly, except for linear
systems, the evolution of quantum systems in phase space (in the
Wigner function picture, for example \cite{Wig,KiNo,Gar}), follows the classical
orbits at best only approximately.

Omn\`es has made some interesting progress in this area \cite{Omn1,Omn2}. He
characterizes a phase space state localized in a region $\Gamma$ of phase space
by a quasi-projector of
the form
\beq
P_{\Gamma} = \int_{\Gamma} dp dq | \psi_{pq} \rangle \langle \psi_{pq} |
\label{1.1}
\eeq
where $ | \psi_{pq} \rangle $ are a set of Gaussian states localized in
$p$ and $q$ (with the coherent states as a special case). These operators
are not exact projectors, however, since the defining property of a projection operator
\beq
P^2 = P
\eeq
is satisfied only approximately by $P_{\Gamma}$,
due to the fact that the states $  | \psi_{pq} \rangle $
are only approximately orthogonal. Bounds may be found on the quantity
$ \| P^2_{\Gamma} - P_{\Gamma} \| $ (in a suitably defined norm)
and this bound give the limits to within which $P_{\Gamma}$ may for practical purposes be
regarded as a projector.
The time evolution of these quasi-projectors may be determined
using results of Hagedorn on the evolution of Gaussian states in a general potential \cite{Hag}.
This leads to the interesting result
\beq
e^{ \ih H t} P_{\Gamma} e^{ - \ih H t } \approx P_{\Gamma (-t)}
\label{1.3}
\eeq
with a calculable error \cite{Omn1,Omn2}, where $\Gamma (-t)$ denotes the classical evolution (backwards in time) of the initial phase space cell $\Gamma$.

In brief, these properties of the quasi-projector Eq.(\ref{1.1})
in the quantum case meet classical
expectations up to certain estimable errors.
However, this leaves the question as to whether there is room for improvement,
in terms of making some of these approximations exact, at least in certain
special cases. There are a number of reasons why one might want to do this.

First of all, when one talks about an ``approximation'', one usually has in mind
an exact result which may be approached in some limit. The above results
only become exact in the limit $\hbar \rightarrow 0 $, which is not really
a physically meaningful limit. It might therefore be of interest to find some non-trivial
situations in the quantum theory in which exact projectors exist, with exactly
specifiable time evolution properties. Such exact results would then form a useful
background from which approximations to the more general case could be constructed.

The second reason concerns the decoherent histories approach to quantum theory,
one of the most useful frameworks in which to address emergent classicality
\cite{Omn1,Omn2,GH1,GH2,Gri,Hal2,Hal3}.
Omn\`es has shown that his results above allow one
to define sets of phase space localized histories with the property of
approximate decoherence \cite{Omn1,Omn2}. Exact projectors with exactly defined time evolution
would give exact decoherence, thereby giving some insight into the relationship
between exact and approximate decoherence of histories.

The purpose of this paper is to show how to construct a class of exact projection
operators with phase space localization. For linear systems, their time evolution
is exactly classical.

A similar task was attempted in a previous paper \cite{HalPQ}. There, inspired by an old idea
of von Neumann \cite{Von}, operators $\hat X, \hat P$ were sought which are in a certain
sense close to the usual position and momentum operators, $\hat x$, $\hat p$,
but which commute, $ [ \hat X, \hat P]  = 0$. From the spectrum of these operators
it is possible to construct a set of exact projection operators $P_{\Gamma_i}$,
with phase space localization, satisfying
\beq
\sum_i P_{\Gamma_i} =1,  \ \ \ \ P_{\Gamma_i} P_{\Gamma_j} = \delta_{ij} P_{\Gamma_i}
\eeq
where $\Gamma_i$ denotes a set of rectangular phase space cells covering the entire phase
space. (This is in turn related to older work on the orthogonalization of the
coherent states \cite{BGZ,Zak1,Zak2}).
Although this work achieved to some extent the goals set out here, it
suffers from some weaknesses. The states used in the construction of the
projection operators were quite complicated. More significantly, the whole
construction only works for rectangular (or piecewise rectangular) cells
that are quite large compared to a quantum cell of size $\hbar$. In addition,
the initial density operator of the system must be coarse-grained over cells
of about the same size. For example, cells of size $10^6 \hbar$ gave good results,
and are comfortably adequate for discussions of the emergent classicality of
macroscopic systems, but one wonders whether it would be possible to find a construction
with much smaller phase space cells.

Here we achieve precisely that. We find exact projectors requiring considerably
less coarse graining of the initial state to give sensible results. This improvement comes
with a price, namely that we can construct a set of just two projectors, $P_{\Gamma}$
and its complement $ \bar P_{\Gamma} = 1 - P_{\Gamma} $, not the infinite set
$P_{\Gamma_i}$ obtained in Ref.\cite{HalPQ}. However, this is in fact sufficient for
discussions of emergent classicality. Furthermore, the properties of these
projectors are considerably simpler under time evolution. In the remainder of this paper we set $\hbar =1 $ for convenience.

\section{Exact Projectors for Circular Phase Space Regions}

\subsection{The Strategy}

How do we modify the quasi-projector Eq.(\ref{1.1}) into an exact projection operator?
This object is a positive hermitian operator and may therefore be diagonalized,
\beq
P_{\Gamma} = \sum_n \lambda_n | \psi_n \rangle \langle \psi_n |
\eeq
where the eigenstates $ | \psi_n \rangle $ are orthogonal and $\lambda_n \ge 0 $.
From the properties of Eq.(\ref{1.1}) one can anticipate that the eigenstates
$ | \psi_n \rangle $ will be phase space localized states and, more importantly,
since $P_{\Gamma}$ is almost a projector, most of its eigenvalues will be close to
$1$ or $0$. This suggests the following strategy for constructing an exact projector:
replace the eigenvalues $\lambda_n$ with eigenvalues $ \tilde \lambda_n$
that are exactly $1$ or $0$. The resulting object,
\beq
E_{\Gamma} = \sum_n \tilde \lambda_n | \psi_n \rangle \langle \psi_n |
\eeq
will be an exact projector which we expect will have
the desired phase space localization properties. (There will clearly be some ambiguity in this procedure for the eigenvalues in the transitional region from $0$ to $1$ but it seems unlikely this ambiguity will be significant).

\subsection{Circular Phase Space Regions}

It is unlikely that this strategy can be carried out in general, for arbitrary
regions $\Gamma$. However, it can be carried out explicitly for circular
regions in phase space with the Gaussian functions $ |\psi_{pq}\rangle $ in Eq.(\ref{1.1})
taken to be the standard coherent states $|z \rangle$. We take these coherent states to be those of
the Hamiltonian
\beq
K = \half (p^2 + q^2)
\label{2.3a}
\eeq
where $z = (q+ i p) / \sqrt{2}$ and
we consider a quasi-projector
\beq
P_{\Gamma} = \int_{\Gamma} \frac {d^2 z} { \pi} | z \rangle \langle z |
\label{2.2}
\eeq
where $\Gamma$ is a circular region of radius $R$ centred on the origin.
We show that this quasi-projector is in fact diagonal in the eigenstates $ | n \rangle$
of the Hamiltonian $K$.
Using $ z = r e^{ i \theta} $ and using the fact that
\beq
\langle z | n \rangle = \frac {z^n} { (n!)^{\half} } e^{ - | z |^2 / 2}
\eeq
we have
\bea
P_{\Gamma} &=& \int_0^R r dr \int_0^{2 \pi} \frac {d \theta} {\pi} | z \rangle \langle z |
\\
&=& \sum_{n,m}  \int_0^R r dr \int_0^{2 \pi} \frac {d \theta} {\pi} \ \frac { (z^*)^n z^m}
{ (n!)^{\half} (m!)^{\half} }\  e^{- | z |^2 } \ | n \rangle \langle m |
\eea
The crucial result is that the $\theta$ integral produces a Kronecker delta,
\beq
\int_0^{2 \pi} \frac {d \theta} {\pi}  e^{ i (m-n) \theta} = 2 \delta_{nm}
\eeq
so we have
\beq
P_{\Gamma} = \sum_n  \lambda_n | n \rangle \langle n |
\eeq
where
\beq
\lambda_n = \frac {2} {n!} \int_0^R dr \ r^{2n+1}  \ e^{-r^2}
\eeq
The dominant contribution to the $r$ integral comes from $ r = n^{\half}$
and it is easily seen that $\lambda_n \approx 0 $ for $ R \ll n^{\half}$
and $ \lambda_n \approx 1 $ for $ R \gg n^{\half}$. The oscillator eigenstates
$ | n \rangle $ are not phase space localized in the same sense as coherent
states, for example, but it is well-known
that the eigenstate $ | n \rangle$
is concentrated inside the circle of radius $n^{\half}$, that is, inside the
classical orbit corresponding to that value of energy. (This will be discussed more below).

The underlying reason that the quasi-projector is diagonal in oscillator
eigenstates in this case is easy to see. Eq.(\ref{2.2}) clearly
satisfies
\beq
e^{ i K s } P_{\Gamma} e^{ - i K s} = P_{\Gamma}
\eeq
since $K$ simply shifts the coherent states along the classical orbits,
which are circles in this case, and this maps the domain of integration $\Gamma$
into itself. So $K$ commutes with $P_{\Gamma}$ and they may be diagonalized in
the same basis.

We thus see that the diagonalization of the quasi-projector is carried out very easily
in this case. Following the general strategy, a simple modification of the
quasi-projector leads to the exact projector
\beq
E_{\Gamma} = \sum_{n=0}^N | n \rangle \langle n |
\label{2.11}
\eeq
We anticipate that this object is localized about a circular region in phase space
of radius $N^{\half}$ centred about the origin, so $N$ needs to be chosen so that
$N^{\half} = R $ for Eq.(\ref{2.11}) to approximate Eq.(\ref{2.2}).

\subsection{Phase Space Localization Properties}

We now demonstrate the localization properties of Eq.(\ref{2.11}) in more detail.
Consider the probability associated with the phase space region $\Gamma$,
as defined using the exact projector Eq.(\ref{2.11}). It may be written in terms of the Wigner functions
of $E_\Gamma$ and the state $\rho$ as
\bea
p( \Gamma ) &=& {\rm Tr} \left( E_{\Gamma} \rho \right)
\nonumber \\
&=& 2 \pi \hbar \int dp dq \ W_E (p,q) W_{\rho} (p,q)
\eea
The Wigner function of $E_{\Gamma}$ is easily obtained from the Wigner functions of the harmonic
oscillator states. It is,
\beq
W_E (p,q) = \sum_{n=0}^N \frac {(-1)^n} { \pi} L_n (r^2 ) e^{ - r^2 /2}
\eeq
where $ r^2 = 2 (p^2 + q^2) $ and $L_n $ are the Laguerre polynomials \cite{Wig}.
This will be non-zero and rapidly oscillating for small $r$. For large $r$
the $L_n(r^2)$ goes like $r^{2n}$ and each term in the summand decays
exponentially for
$r \gg n^\half $. It follows that $W_E(p,q)$ will be small for
$ r \gg (2N)^{\half}$.

By contrast, the Omn\`es quasi-projector, Eq.(\ref{1.1}), will have a Wigner function
that takes an approximately constant positive value for small $r$, decaying to
zero for large $r$. Hence the price of an exact projector versus a quasi projector
is the rapid oscillations, which may mean that the exact projector does not have good
localization properties for certain types of states.

It is also enlightening to look at the probability $p(\Gamma)$ using
the $P$-function representation \cite{Gar} of the density operator,
\beq
\rho = \int d^2z \ P(z) | z \rangle \langle z |
\eeq
We then have
\beq
p(\Gamma) = \int d^2z \ P(z) \ \langle z | E_{\Gamma} | z \rangle
\eeq
where
\bea
\langle z | E_{\Gamma} | z \rangle &=& \sum_{n=0}^N | \langle n | z \rangle|^2
\nonumber \\
&=& \sum_{n=0}^N \frac{ |z|^{2n} } { n!} e^{- | z |^2 }
\eea

This is approximately $1$ for $ | z |^2  \ll N $ and approximately zero for
$ | z |^2 \gg N $. Hence the probability is
\beq
p( \Gamma ) \approx \int_{ | z |^2 < N} d^2z \ P(z)
\label{2.17}
\eeq
This probability will be physically sensible as long as the $P$-function
is positive. If it is negative, positive and negative parts of $P$ may cancel
in the integration yielding an unrepresentatively small value for the probability for certain
regions of phase space.  It is well-known that coarse graining of the $P$ function over phase space regions of size order $\hbar$ make it positive \cite{DiKi}, which suggests that Eq.(\ref{2.17}) will therefore give good phase space probabilities as long as the region integrated over is larger than a few Planck-sized cells. This therefore represents a big improvement on the earlier work Ref.\cite{HalPQ} which required coarse graining over a large number of Planck cells.

\subsection{More General Phase Space Regions}

The construction Eq.(\ref{2.11}) yields an exact projector localized around a circle in phase space centred around the origin. This is simply generalized in a number of ways.
To obtain a projector centred about another point $p, q$ in phase space,
we apply the unitary shift operator
\beq
U(p, q) = \exp(  i p \hat q - i q \hat p  )
\label{3.1}
\eeq
The projector
\beq
E_{{pq}} = U(p, q) E_{\Gamma} U^{\dag} (p, q)
\label{3.2}
\eeq
is then localized to a circle centred around the point $(p, q)$. One may also apply standard unitary operators
to $E_{pq}$ which produce squeezing and rotations in phase space, turning the circular regions into
ellipses of any eccentricity and orientation \cite{KiNo,Gar}. Hence it is possible to construct an exact projector localized about any elliptical region in phase space. 

Exact projectors localized to general elliptical regions could be useful even if the dynamics generates phase space cells of more complicated shapes -- one can use the projector to ask if the system is within the smallest elliptical region containing the given phase space cell of complicated shape. This could be useful if the boundary is not too irregular. Still, it is of interest to generalize the construction given so far.

For regions $\Gamma$ in phase space of more general shape,
the diagonalization procedure for the projector Eq.(\ref{1.1}) is unlikely
to be implementable in practice.
However, the construction of Eq.(\ref{2.11}) suggests another strategy which may be useful
for some geometries. This is to keep
the same form Eq.(\ref{2.11}) but
to {\it choose} the Hamiltonian operator $K$ used in its construction so that its
eigenstates $ | n \rangle$ are localized in a given region $\Gamma$ of any reasonable shape.
More precisely, suppose that the region $\Gamma$ is a generic region
centred around the origin with boundary $\partial \Gamma$. It will probably
be necessary to require that $\partial \Gamma$ is reasonably smooth.
Clearly $\partial \Gamma $ is a closed curve around the origin, $ (p(s), q(s))$
where $ s $ is periodic. Classically, it seems very plausible that, at least for
some interesting class of boundary curves, we can {\it choose}
the potential $U$ in a Hamiltonian
\beq
K = \frac {p^2} {2} + U (q)
\eeq
so that the boundary curve $\partial \Gamma$ is one of its
integral curves, with fixed energy $K_{\Gamma}$, say. Furthermore,
integral curves with lower values of energy will be closed curves lying inside
$\partial \Gamma$.

When we look at the spectrum of $\hat K$ in the quantum theory, it will
have properties similar to the classical paths. They will also be qualitatively
similar to the spectrum of the simple harmonic oscillator case already considered.
This can be seen, for example, from studies of the Wigner function of the eigenstates
of bound state Hamiltonians \cite{Berry}.
That is, the eigenstates $ | n \rangle $ up to some value $ N \approx 2 K_{\Gamma}$
will be localized in $\Gamma$ and decay rapidly outside. Hence, the projector
of the form Eq.(\ref{2.11}) will have the desired property of localization in
the region $\Gamma$. In essence what we are doing here is distorting the construction for the case
where $K$ is a simple harmonic oscillator Hamiltonian and $\Gamma $ a circular
region, in such a way that the circular region is turned into a general region.

This procedure clearly will not work for any region $\Gamma$ -- the main restriction is that the curves must be reflection-symmetric under $p \rightarrow - p$. Furthermore, this construction has not be spelled out in detail, so is only offered as a suggestion. However, it does indicate that a construction for some more general and interesting geometries is possible.

\section{Time Evolution}

We now consider the time evolution of the exact projectors on circular regions
defined in the previous section. At this point we stress that there are potentially
two Hamiltonian operators to consider. One is the simple harmonic oscillator
Hamiltonian K associated with the coherent states defined above. This is simply an
operator used in the definition of the projectors. The other is the physical Hamiltonian
of the system in question, which we denote $H$, used to evolve the system in time.
These two operators need not be the same. However, we here consider the case in
which $H = K$ and consider the time evolution of the projector Eq.(\ref{3.2}).

We first note that the projector $E$ commutes with $H$, so the time evolution
of $E_{pq}$ is entirely in the unitary shift operators Eq.(\ref{3.1}). We also note
that
\beq
e^{ i H t } U ( p,q) e^{ - i H t } = \exp(  i p \hat q (t) - i q \hat p(t)  )
\label{3.3}
\eeq
where
\bea
\hat q(t) &=& \hat q \cos t  + \hat p \sin t
\label{3.4} \\
\hat p(t) &=& - \hat q \sin t  + \hat p \cos t
\label{3.5}
\eea
It follows that
\beq
e^{ i H t } U ( p,q) e^{ - i H t } = U (p(-t), q(-t) )
\eeq
where $p(t)$ and $q(t)$ are defined in the same way as
Eqs.(\ref{3.4}), (\ref{3.5}). Applying these results to Eq.(\ref{3.2}), we find
\bea
e^{ i H t} E_{pq} e^{ - i H t} &=& U (p(-t), q(-t) ) E_{pq}  U^{\dag} (p(-t), q(-t))
\\ \nonumber
&=& E_{p(-t)q(-t)}
\label{3.5}
\eea
That is, unitary time evolution of a projector centred on the point $p,q$ consists
simply of shifting it (backwards in time) along the classical trajectory
with initial data $p,q$.

This is similar to the appealing earlier result Eq.(\ref{1.3}) except
that this result is exact. It is exact firstly, because the system is linear, but secondly,
because of the special relationship between the system Hamiltonian and the construction of the
projector, which means that the projector commutes with the Hamiltonian (which is not true of the Omn\`es quasi-projectors).

A more general situation to consider is that in which the projectors onto
circular regions are subject to time evolution under a more general Hamiltonian,
so $K$ and $H$ are different. We expect a similar quasiclassical result to
hold. However, classically, a circular region of phase space evolving under
a general Hamiltonian will not remain circular, so it is necessary to consider
the construction of exact projectors onto phase space regions which are not
circular.

\section{Exactly Decoherent Phase Space Histories}

We now briefly describe the connection of the above results to exact decoherence of histories.
In the decoherent histories approach to quantum theory \cite{Omn1,Omn2,GH1,GH2,Gri,Hal2,Hal3}, histories are represented by
projection operators $P_{\alpha}$ at each moment of time, where
$P_{\alpha} P_{\alpha'} = \delta_{\alpha \alpha'} P_{\a}$ and $\sum_{\alpha} P_\alpha = 1$.
It is sufficient for what we are doing here to focus on the simplest non-trivial case of histories
characterized by two moments of time and in this case the candidate probability for a set of histories is
\beq
p(\a_1, \a_2) = {\rm Tr} \left(P_{\a_2} e^{ - i H t} P_{\a_1} \rho P_{\a_1} e^{ i H t} \right)
\eeq
For these to be satisfactory probabilities, the condition of decoherence must be satisfied, which means that
the decoherence functional,
\beq
D(\a_1, \a_1', \a_2) = {\rm Tr} \left(P_{\a_2} e^{ - i H t} P_{\a_1} \rho P_{\a_1'} e^{ i H t} \right)
\eeq
is zero for $\a_1 \ne \a_1'$. This condition is typically satisfied only approximately (which has led to some discussion as to what this actually means).

There are, however, some simple circumstances in which the decoherence condition is satisfied exactly.
The first and obvious one is that in which the projectors commute with $H$, that is, they are histories
of conserved quantities. The second and perhaps more subtle one, is that in which there exists an exact relation of the form,
\beq
e^{ i H t} P_{\a_2} e^{ - i H t} = P_{\a_1 (t)}
\label{4.3}
\eeq
where the label $\a_1 (t)$ depends on $\a_2$ and on $t$. There is clearly exact decoherence in this case, as long as the labels $\a_1$ and $\a_2$ are properly chosen. In simple terms it arise because the quantum theory has a sort of exact determinism, more general than simple conservation.

The phase space projectors considered here provide an example of this sort of determinism and hence of exact decoherence. To see this, we consider the simple harmonic oscillator example above and choose
$P_{\a_1}$ and $P_{\a_2}$ to be exact projections onto circular phase space regions $\Gamma_1$ and $\Gamma_2$ related
by classical evolution (and their complements).
We therefore let $\a_1$ take two values and let the projectors $P_{\a_1}$ denote the projector $E_{pq}$ onto a phase space region $\Gamma_1$, defined above in Eq.(\ref{3.2}), and its negation, $ \bar E_{pq} = 1 - E_{pq}$.
Similarly, we let $P_{\a_2}$ denote the projector onto the time-evolved phase space region $\Gamma_2$,
$E_{p(t)q(t)}$ and its negation, $1 - E_{p(t)q(t)}$.
We immediately see from Eq.(\ref{3.5}) that these projectors satisfy a relation of the form
Eq.(\ref{4.3}) and there is exact decoherence. To see this explicitly, a typical off-diagonal term in the
decoherence function is
\beq
D(\a_1, \a_1', \a_2) = {\rm Tr} \left( e^{ i H t} E_{p(t)q(t)} e^{ - i H t}
E_{ pq} \rho \bar E_{pq} \right)
\eeq
which is clearly zero (from Eq.(\ref{3.5}) with $t$ replaced by $-t$) since the projectors are all exact projectors.

This analysis is easily extended to a set of phase space localized projectors at an arbitrary sequence of times. Given exact decoherence, one then can consider the probabilities for such phase space histories. We therefore consider a sequence of phase space regions $\Gamma_1, \Gamma_2, \cdots \Gamma_n$, which are all related by classical evolution.
It is the easy to see that the probability for a set of histories characterized by projectors of the above type is quite simply $ {\rm Tr} ( E_{pq} \rho) $, since a relation of the form Eq.(\ref{3.5}) holds and the projectors are exact, so a string of such projectors collapses down to just one. In fact, we could
choose the initial state to be the phase space localized state $\rho = E_{pq} / {\rm Tr} (E_{pq})$, in which case the probability of the above sequence of phase space regions is then exactly $1$, so there is an exact classical determinism in this model.

These results are an exact version, in the special case of a simple harmonic oscillator, of the approximate decoherence
discussed by Omn\`es \cite{Omn1,Omn2}, using the approximate relation Eq.(\ref{1.3}).

\section{Summary and Conclusions}

We have shown how to construct exact projection operators which are localized about regions of phase space.
The construction was given and explored in detail for circular regions and easily extends to elliptical regions of arbitrary centre, eccentricity and orientation using unitary transformations. We sketched how the construction might proceed for more general regions. The time evolution of projections onto circular regions was considered. It is particularly simple and was used to give an example of exactly decoherent sets of phase space histories, with an exact classical determinism.

\section{Acknowledgements}

I am grateful to James Yearsley for useful conversations and for preparing the figure.

\bibliography{apssamp}

\end{document}